\documentclass[a4paper,showkeys,floatfix,aps,pre,reprint,groupedaddress]{revtex4-1}
\usepackage{color}
\usepackage{amsmath,amssymb}
\usepackage{graphics,graphicx}
\usepackage{dcolumn,bm}
\usepackage{psfrag}
\usepackage{soul}
\usepackage{enumitem}

\begin{document}

\title{Creep-like behavior in athermal threshold dynamics: Effects of disorder and stress}

\author{Subhadeep Roy}
\email{sroy@eri.u-tokyo.ac.jp}
\author{Takahiro Hatano}
\email{hatano@eri.u-tokyo.ac.jp}
\affiliation{Earthquake Research Institute, University of Tokyo, 1-1-1 Yayoi, Bunkyo, 113-0032 Tokyo, Japan.}

\date{\today}

\begin{abstract}
\noindent 
We study the dynamical aspects of a statistical-mechanical model for fracture of heterogeneous media: the fiber bundle model with various interaction range.
Although the model does not include any thermal activation process, the system exhibits creep-like behaviors under a constant load being slightly above the critical value.
These creep-like behaviors comprise three stages: in the primary and tertiary stages, the strain rate exhibits power-law behaviors with time, which are well described by the Omori-Utsu and the inverse Omori laws, respectively, although the exponents are larger than those typically observed in experiments. A characteristic time that defines the onset of power-law behavior in the Omori-Utsu law is found to decrease with the strength of disorder in the system. The analytical solution, which agrees with the above numerical results, is obtained for the mean-field limit.
Beyond the mean-field limit, the exponent for the Omori-Utsu law tends to be even larger but decreases with the disorder in the system.
Increasing the spatial range of interactions, this exponent is found to be independent of disorder and to converge to the mean-field value.
In contrast, the inverse Omori law remains independent of the spatial range of interaction and the disorder strength.
\end{abstract}


\maketitle


\section{Introduction}
The two major laws for the earthquake statistics are the Omori-Utsu law \cite{OmoriMain1,Utsu} and the Gutenberg-Richter (GR) law \cite{GR}.
The latter describes the frequency of earthquakes with respect to their magnitude and the former describes the rate of aftershocks decreasing in a power-law fashion with the time elapsed from a mainshock:
\begin{align}\label{intro1}
n(t)=\displaystyle\frac{k}{(t+c)^p},
\end{align}
where $n(t)$ is the aftershock rate as a function of the elapsed time $t$ from a mainshock, $p$ and $k$ the positive constants, and $c$ the time constant. The exponent $p$ varies from $0.7-1.6$ \cite{Utsu,OmoriReview}, while the $c$-value may depend on the mainshock magnitude and a magnitude cutoff for aftershocks \cite{Omori1}. Generally, the precise estimate of $c$-value is difficult because it is strongly affected by the detection ability of aftershocks, which is degraded by the mainshock coda. Nevertheless, careful analyses have revealed that the $c$-value takes a definite non-zero value \cite{Peng,Enescu} and exhibits a systematic dependence on the faulting geometry \cite{Narteau}. As the faulting geometry correlates with the differential stress on faults, it also implies the stress dependence of the $c$-value: it decreases for larger stress \cite{Narteau}.

Interestingly, the statistics for micro-fracture events in the laboratory scale shares many aspects with those for earthquakes: the GR \cite{Scholz} and the Omori-Utsu laws \cite{Hirata,Schubnel}. Moreover, in creep tests, in which the constant stress is applied to a specimen, the strain rate decreases in a power-law fashion resembling the Omori-Utsu law \cite{Andrade}. This stage is referred to as the primary creep and it is followed by the secondary creep with nearly time-independent strain rate. In the subsequent tertiary creep, the strain rate increases in a power-law fashion, which leads to breakdown of a specimen. This power-law acceleration of the strain rate is described by the inverse Omori law:
\begin{equation}
\dot{\epsilon}(t) \propto (t_f - t+c')^{-p'},
\end{equation}
where $t_f$ is the time of breakdown, $c'$ the time constant, and $p'$ the positive exponent. However, the inverse Omori law is not usually observed for earthquakes \cite{Bouchon}, whereas it is common in material failure. In their original form, the Omori-Utsu and the inverse Omori laws describe the rate of micro-fracture events but they can be reinterpreted in terms of the strain rate. In this study, we refer to these laws in terms of the strain rate.

There have been some simple models that can reproduce these power-law behaviors.
Most of them are classified into the fiber bundle model \cite{Daniels,RevModPhys82,Book}.
This is an assembly of fictitious fibers that support the mechanical load in parallel.
Each fiber has its own failure threshold, which is randomly set according to a specific probability distribution function.
Aiming at reproducing creep-like behaviors, some studies adopt probabilistic rules for the elementary failure process, which may model thermal activation processes \cite{Ciliberto,Pradhan2003,Shcherbakov,Ben-Zion,Saichev} or introduce additional variables that may correspond to the accumulated damage in the fibers \cite{Danku}. These attempts, which are regarded as the extension of the original model \cite{Daniels}, may be legitimate because creep involves thermal activation processes as its microscopic origin.
In contrast, however, Pradhan and Hemmer adopted a simple deterministic model to show that it is sufficient to reproduce creep-like behaviors qualitatively \cite{Pradhan2007}. On the other hand, they did not discuss the power-law behaviors such as the Omori-Utsu and the inverse Omori laws. Additionally, their analysis is limited to the mean-field limit and a particular strength of disorder.

In the present study, we investigate the time evolution toward breakdown in a simple fiber bundle model under a constant load for both the mean field and the local stress concentration cases. Although the model does not include any thermal activation process, it resembles most properties observed in creep tests including the above-mentioned three stages. Particularly, the Omori-Utsu and the inverse Omori laws are reproduced and the exponent $p$ and the $c$-value are obtained. We show the dependence of $c$-value on the external load and disorder in the system.

In the next section, we provide a brief discussion of the model followed by the analytical results given in section III. The numerical results for the mean field limit as well as the local load sharing model are discussed in details in section IV. In the final section we have discussed our findings and summarized the chances of future works.


\section{Description of the model}
Here we adopt a fiber bundle model as in the previous studies. Initially, the $L$ intact fibers support the load $F$ in parallel, resulting in the stress of $f=F/L$ on each single fiber. Each fiber has its own fracture strength chosen randomly from a certain distribution. The dispersion of the fracture strength characterizes the disorder in the model. If the applied stress exceeds any of the threshold values, the corresponding fibers break irreversibly. After a rupture event, the load that has been supported by the broken fibers is redistributed among the remaining intact fibers. In this literature, two kinds of redistribution models are commonly used: (i) the global load sharing (GLS) model, in which the load is redistributed equally among all the other surviving fibers \cite{Pierce,Daniels} and (ii) the local load sharing (LLS) model, in which the load is redistributed only to the surviving neighbors \cite{Phoenix,Smith,Newman,Harlow2,Harlow3,Smith2}. The GLS model is regarded as the mean-field model as the range of load redistribution is infinite.

In both the models, the stress on the unbroken fibers increases upon redistribution of the load and therefore a single rupture event can cause a cascade of further rupture events. At a given load of $F$, the system eventually comes to a stable state after a cascade of ruptures, otherwise all the fibers fail. In the former case, the load is increased slightly for a system to reach another stable state with less surviving fibers at larger stress. Such an increment process may be repeated until all the fibers break after a series of cascades. The repetition of the force increments enables one to define the critical load $F_c$, at which all the fibers are broken; namely, there are no stable states above the critical stress $F_c$. The value of $F_c$ depends on the disorder in the system \cite{sroyepl}, described by dispersion in the strength of fibers . Note that it also depends on the system size $L$ in the local load sharing limit \cite{Pradhan2003,llssystemsize1,sroy}.

In this study, we investigate the time evolution of the model at a constant load, which is slightly above the critical value. The system is eventually led to the breakdown under such a large load but we can still investigate the dynamics towards breakdown by introducing the relaxation time of the load redistribution. Namely, the breaking of fibers and the following load redistribution should take the finite relaxation time $\tau$, which can be regarded as a single time step. Note that this time constant is assumed to be zero in a conventional algorithm \cite{Daniels}. 

Here we adopt the following algorithm: the total load remains at $F(\approx F_c)$ throughout the time evolution. The initial stress is thus $F/L$ at $t=0$. Then the fibers having the strength lower than $F/L$ should break and as a result the load is redistributed to all the remaining fibers (the GLS model) or only to the neighbors of the broken fibers (the LLS model). In any case, due to the load redistribution, some fibers are overloaded beyond their strength, resulting in further ruptures at the next time step, $t=\tau$. This defines a single time step in our algorithm and is repeated until all the fibers are broken. This relaxation time $\tau$ may depend on any physical ingredients such as the stress, time or the temperature. Here we regard $\tau$ as a constant for simplicity. Then the model is essentially the same as that investigated by Pradhan and Hemmer \cite{Pradhan2007}. They investigated only the GLS but the dynamics of the system depends largely on the nature of load redistribution. In this paper we investigate the time evolution in both models: the GLS and the LLS.


\section{Analysis on Global load sharing model}
In this section the dynamics of the above mentioned GLS model, which is the mean field limit, is studied analytically. Note that the stress is identical for all the fibers in the GLS model. This allows one to discuss the system behavior analytically for some simple threshold distributions.
 
Writing the threshold distribution as $p(y)$, a general expression for the number of remaining fibers after the $i^{th} (i=1, 2, \cdots)$
redistribution is
\begin{equation}
L_{i} = L_{0} -  \int_0^{f_{i-1}} L_0 p(y)dy.
\end{equation}
where $L_{0}$ is the initial number of fibers, $L_{i}$ the number survived after the  $i^{th}$ redistribution, and $f_{i-1}$ the force per fiber at the previous time step $i-1$.
This can be rewritten in terms of fraction $n_i = L_i/L_0$
\begin{equation}
n_{i} = 1 -  \int_0^{f_{i-1}} p(y)dy.
\end{equation}
Using $n_i = f_0/f_i$, where $f_0$ being the strain at $t=0$, the above equation is rewritten in terms of $f$.
\begin{equation}
\label{recursive}
f_{i} = \frac{f_0}{1 -  \int_0^{f_{i-1}} p(y)dy} = \frac{f_0}{\int_{f_{i-1}}^{\infty} p(y)dy}
\end{equation}
This is the recursive relation for $f$.
One can also consider a differential equation by using
$(f_{i+1}-f_{i})/\tau \simeq \dot{f}$, where $\tau$ is the duration of one time step, $i$ to $i+1$:
\begin{equation}
\label{ODE}
\tau\dot{f} = \frac{f_0}{1 -  \int_0^{f} p(y)dy} - f.
\end{equation}
Therefore, the time evolution of the present system is solely determined by the threshold distribution $p(f)$ and the initial condition $f_0$.
As $f_0=F/L_0$, choosing $f_0$ is identical to determine the external load $F$.
Note that $f$ should be proportional to the strain of the system as the elastic modulus is supposed to be identical to all the fibers.
Therefore, $\dot{f}$ should be proportional to the strain rate.

\subsection{Uniform threshold distribution}
For a uniform threshold distribution defined in $[f_{\rm max}-\delta, f_{\rm max}]$, the integral in Eq. (\ref{ODE}) is easily solved to give
\begin{equation}
\tau \dot{f} = \frac{f_0\delta}{f_{\rm max} - f} -f.
\end{equation}
This is rewritten as
\begin{equation}
\tau \dot{f} =\frac{\left( f-\frac{f_{\rm max}}{2}\right)^2+f_0\delta-\frac{f_{\rm max}^2}{4}}{f_{\rm max} - f}.
\end{equation}
This is more simplified as
\begin{equation}
\label{nondim}
\tau \dot{x} =\frac{\left( x-\frac{1}{2}\right)^2+\zeta}{1-x},
\end{equation}
where
\begin{eqnarray}
x  &:=& \frac{f}{f_{\rm max}}\\
\label{epsilon}
\zeta &:=& \frac{f_0\delta}{f_{\rm max}^2}-\frac{1}{4}.
\end{eqnarray}
By choosing $\tau$ as the time unit, we realize that there is only one non-dimensional parameter $\zeta$ that controls the time evolution of $x$.

\begin{itemize}
\item If $\zeta$ is non-positive, Eq. (\ref{nondim}) has steady-state solutions of 
$x=1/2\pm \sqrt{-\zeta}$: $x=1/2 - \sqrt{-\zeta}$ is the stable fix point
and the other is the unstable fix point.
Starting from the initial condition $x_0 < 1/2 - \sqrt{-\zeta}$, the system relaxes to the stable fixed point exponentially.
Although the time derivative of $x$ is negative for $x$ between these two fix points, it should be interpreted as $\dot{f}=0$ because the system is essentially irreversible.

\item At $\zeta=0$, the saddle-node bifurcation occur.
Namely, these two fixed points merge together and annihilate.
This bifurcation is actually present for more general threshold distributions, and therefore they yield common behaviors near the bifurcation point.

\item For positive $\zeta$, there is no fixed point and the system undergoes breakdown.
The exact solution of Eq. (\ref{nondim}) is given as
\begin{align}
\label{solution_uniform}
t_m-t = &\frac{1}{2}\log\left[\left(\frac{1}{2}-x\right)^2 + \zeta\right] \nonumber \\
&+\frac{1}{2\sqrt{\zeta}} \tan^{-1}\left(\frac{1/2-x}{\sqrt{\zeta}}\right),
\end{align}
where $t_m$ is an integral constant.
Here we consider a system close to the bifurcation point, $\zeta \ll 1$.
Then the first term is negligible and one gets the following expression.
\begin{equation}
\label{solution_uniform_tan}
x \simeq \frac{1}{2} + \sqrt{\zeta}\tan[2\sqrt{\zeta}(t-t_m)].
\end{equation}
Because the above equation should give $x=x_0$ at $t=0$, 
\begin{equation}
t_m \simeq \frac{1}{2\sqrt{\zeta}}\tan^{-1} \left(\frac{1/2-x_0}{\sqrt{\zeta}}\right).
\end{equation}

The time evolution of the system is fully described by Eqs. (\ref{solution_uniform}) or (\ref{solution_uniform_tan}).
A practically important quantity is the time of breakdown, $t_f$, where the surviving fibers vanish: Namely, the force per fiber diverges.
Considering Eq. (\ref{solution_uniform}), the time of the breakdown is given by
\begin{equation}
\label{tf}
2\sqrt{\zeta}(t_f-t_m)=\pi/2.
\end{equation}
\end{itemize}
Importantly, Eq. (\ref{solution_uniform}) implies both the Omori-Utsu and inverse Omori laws for the primary and the tertiary stages, respectively.\\

\textit{\textbf{The Omori law}} :
The primary stage is characterized by $x_0 < 1/2$.
In this case $(1/2-x_0)/\sqrt{\zeta} \gg 1$ and therefore $t_m\simeq \pi/4\sqrt{\zeta}$.
We can thus write
\begin{equation}
\label{expand_a}
2\sqrt{\zeta} t_m = \frac{\pi}{2} - f(x_0),
\end{equation}
where $f(x_0) > 0$.
Inserting Eq. (\ref{expand_a}) into Eq. (\ref{solution_uniform}), 
\begin{equation}
\label{solution_uniform2}
x \simeq \frac{1}{2} - \frac{\sqrt{\zeta}}{\tan[2\sqrt{\zeta}t+f(x_0)]}
\simeq \frac{1}{2} - \frac{1}{2t+f(x_0)/\sqrt{\zeta}}
\end{equation}
for small $t$.
Taking the initial condition into account, this leads to
\begin{equation}\label{ref1}
x \simeq \frac{1}{2} - \frac{1}{2[t+1/(1-2x_0)]}.
\end{equation}
Therefore,
\begin{equation}\label{MainOmori}
\dot{x} \simeq \frac{1}{2[t+1/(1-2x_0)]^2}.
\end{equation}
This is the Omori law with $p=2$ and 
\begin{equation}\label{c-value}
c=\displaystyle\frac{1}{1-2x_0}.
\end{equation}
We are thus led to the concrete expression for the $c$-values.\\

\textit{\textbf{The inverse Omori law}} : 
Inserting Eq. (\ref{tf}) into Eq. (\ref{solution_uniform}) and rewriting the time $t$ as $t= t_f - t'$, one can get
\begin{equation}
x \simeq \frac{1}{2} + \sqrt{\zeta}\tan[\frac{\pi}{2}-2\sqrt{\zeta}t']
\simeq \frac{1}{2} + \frac{1}{2(t_f -t)}.
\end{equation}
This leads to the accelerating creep in the tertiary regime.
\begin{equation}\label{MainInvOmori}
\dot{x}\simeq \frac{1}{2(t_f -t )^2}.
\end{equation}
Note that the $c$-value is not visible in this tertiary stage, whereas a non-zero $c$-value is obtained in the primary stage.


\subsection{General relation between saddle-node bifurcation and the power-law behaviors}
We discuss more general threshold distributions taking advantage of saddle-node bifurcation.
First we discuss the nature of fixed points, which are the solutions of the following equation.
\begin{eqnarray}
\label{fixedpoints}
f &=& \Phi(f),\\
\label{Phi}
\Phi(f) &=& \frac{f_0}{\int_{f}^{\infty} p(y)dy}.
\end{eqnarray}
Since $p(y)$ is positive, $\Phi(f)$ is a monotonically increasing function of $f$.
Because $\Phi(0) = f_0 > 0$, Eq. (\ref{fixedpoints}) may have some solutions.

\begin{figure}[ht]
\centering
\includegraphics[width=7cm, keepaspectratio]{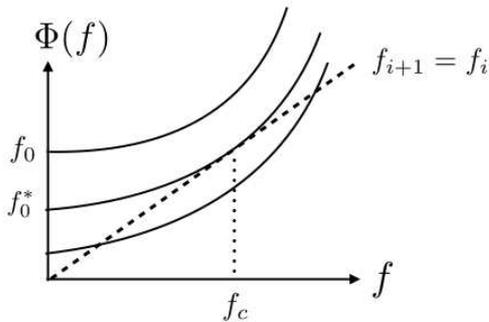}
\caption{Variation of the function $\Phi(f)$ with increasing force per fiber, $f$. 
The time evolution is given by $f_{i+1} = \Phi(f_i)$ and $f_0$ is the initial condition for $f$.
Note that $\Phi(0)=f_0$.
The plots are shown for $f_0<f_0^{\ast}$, $f_0=f_0^{\ast}$ and $f_0>f_0^{\ast}$ (from the bottom to the top), respectively.}
\label{schematic}
\end{figure} 

Let us suppose that saddle-node bifurcation occurs as a critical initial stress, $f_0=f_0^*$.
Then, because $\Phi(f)$ is tangent to $f$ at the bifurcation point, one can consider an expansion of $\Phi(f)$ around $f=f_c$.
\begin{equation}
\label{expansion}
\Phi(f) \simeq f_c + (f-f_c) + a (f-f_c)^2 +\cdots ,
\end{equation}
where $a := \frac{1}{2}\partial^2\Phi/\partial f^2|_{f_c}$ is assumed to be positive.
If $f_0$ is only slightly larger than $f_0^*$, one may write 
\begin{equation}
\label{expansion2}
\Phi(f) \simeq \zeta + f_c + (f-f_c) + a (f-f_c)^2 +\cdots ,
\end{equation}
where $\zeta > 0$.
Truncating the above expansion at the second order, one can write an approximate time evolution equation.
\begin{equation}
\label{truncation}
\tau\dot{f} = \Phi(f) - f \simeq \zeta + a (f-f_c)^2.
\end{equation}
This equation is integrated by separating the variables and gives
\begin{equation}
\sqrt{a\zeta}(t-t_1) = \arctan\left[\sqrt{\frac{a}{\zeta}}(f -f_c)\right]
- \arctan\left[\sqrt{\frac{a}{\zeta}}(f_1-f_c)\right],
\end{equation}
where $f$ and $f_1$ denote $f(t)$ and $f(t_1)$, respectively.
Choosing $t_1=t_m$ and $f(t_m)=f_c$, this equation reduces to
\begin{equation}
\label{solution_general}
f(t) = f_c + \sqrt{\frac{\zeta}{a}}\tan\left[\sqrt{a\zeta}(t-t_m) \right],
\end{equation}
which is identical to Eq. (\ref{solution_uniform}).
Therefore, one can obtain the Omori-Utsu and the inverse Omori laws in the same manner as shown in the previous subsection.
Particularly,
\begin{equation}
\dot{f} = \frac{\tau/a}{\left[ t + \tau/a (f_c-f_0) \right]^2}.
\end{equation}
Therefore, the exponent $2$ should be robust for a wide class of systems that undergoes saddle-node bifurcation.
The time constant $c$ for the Omori law is given by
\begin{equation}
\label{c-value_general}
c = \frac{\tau}{a (f_c-f_0)}.
\end{equation}

In the above discussion, the positiveness of $a$ is crucial.
Note also that these power-law behaviors are realized only in a finite range of $f$ where the expansion of Eq. (\ref{expansion}) can be truncated at the second order.
Namely, Eq. (\ref{truncation}) must hold in a sufficiently wide range of $f$ for the realization of power-law behaviors.
This implies that the saddle-node bifurcation itself is not a sufficient condition for the power-law behaviors.
This valid range of the quadratic approximation and the sign of $a$ depend on the detail of $p(f)$, and therefore we discuss this for some examples in the next subsection.
Importantly, in some special cases the inverse Omori law can be observed without saddle-node bifurcation.
Therefore, the bifurcation is indeed not a necessary condition for a power-law behavior.


\subsection{Power-law distribution}\label{Powerlaw_Analytical}
As the uniform threshold distribution may be a little bit artificial, one should consider some other threshold distributions.
Among them, the power law distribution is particularly instructive as the system exhibits more complex behaviors than the uniform distribution case.
It is also important in view of the geophysical systems as the heterogeneities in solid earth systems are often fractal.

The threshold distribution $p(f)$ is proportional to $f^{-\alpha}$ ($\alpha>0$) within the range of $[f_{\min}, f_{\max}]$ and vanishes otherwise.
For $\alpha\neq 1$, the distribution reads
\begin{equation}
p(f) = \frac{1-\alpha}{f_{\max}^{1-\alpha}- f_{\min}^{1-\alpha}}f^{-\alpha}.
\end{equation}
This leads to
\begin{equation}
\Phi (f) = \left\{
\begin{array}{l}
\frac{f_{\min}^{1-\alpha}- f_{\max}^{1-\alpha}}{f^{1-\alpha}- f_{\max}^{1-\alpha}} f_0,
 \ (f \ge f_{\min})\\
 \\
f_0. \hspace{2cm} (f  < f_{\rm min})
\end{array}
\right.
\end{equation}
Then the time evolution equation is given by
\begin{equation}
\label{TE_powerlaw}
\dot{f} = \frac{f_{\min}^{1-\alpha}- f_{\max}^{1-\alpha}}{f^{1-\alpha}- f_{\max}^{1-\alpha}} f_0 - f,
\end{equation}
and therefore the fixed points are the solution of 
\begin{equation}
\label{FP_powerlaw}
f (f^{1-\alpha}- f_{\max}^{1-\alpha}) = f_0 (f_{\min}^{1-\alpha}- f_{\max}^{1-\alpha}),
\end{equation}
where $ f_{\min}\le f <  f_{\max}$.
Noting that the right hand side of the above equation is a constant, the nature of the fixed points depends on the behavior of the left hand side, which largely depends on the exponent $\alpha$.
As explained below, the saddle-node bifurcation occurs only for $\alpha <2$, and therefore the power-law behavior is not expected for $\alpha \ge 2$.
Nevertheless, the inverse Omori law is observed at $\alpha=2$.
This illustrates that the bifurcation is not a necessary condition of the power-law behaviors.

\begin{itemize}
\item For $\alpha>2$, the left hand side of Eq. (\ref{FP_powerlaw}) is a monotonically decreasing function from the infinity to the negative infinity as $f$ varies from zero to the infinity. 
Therefore, in view of Eq. (\ref{TE_powerlaw}), there must be one unstable fixed point.
In this case the system fails quickly starting from $f_0$ that is larger than a critical value, $f_{0}^*$.
This is given by inserting $f=f_0=f_{0}^*$ in Eq. (\ref{FP_powerlaw}).
Therefore, the power law behavior is not observed for $\alpha>2$.

\item For $\alpha<2$, the left hand side of Eq. (\ref{FP_powerlaw}) is a concave function of $f$ for $0<\alpha<1$, or convex for $1<\alpha<2$.
Therefore, there must be two fixed points at sufficiently small $f_0$, and they should merge at a critical value of $f_0$.
Namely, saddle-node bifurcation occurs.
This bifurcation point is determined by combining Eq. (\ref{FP_powerlaw}) and 
\begin{equation}
\label{CP_powerlaw}
\Phi'(f) = (\alpha-1)\frac{f_{\min}^{1-\alpha}- f_{\max}^{1-\alpha}}{(f^{1-\alpha}- f_{\max}^{1-\alpha})^2} f_0 f^{-\alpha}=1.
\end{equation}
Equations (\ref{FP_powerlaw}) and (\ref{CP_powerlaw}) lead to
\begin{equation}
f (f^{1-\alpha}- f_{\max}^{1-\alpha})\left[2-\alpha-\left(\frac{f}{f_{\max}}\right)^{\alpha-1}\right]=0,
\end{equation}
which gives
\begin{equation}
f = (2-\alpha)^{1/(\alpha-1)}f_{\max}.
\end{equation}
The critical initial condition is given by inserting the above equation to Eq. (\ref{FP_powerlaw}).
\begin{equation}
f_0 = \frac{(\alpha-1)  (2-\alpha)^{(2-\alpha)/(\alpha-1)}}{f_{\min}^{1-\alpha}- f_{\max}^{1-\alpha}}f_{\max}^{2-\alpha}.
\end{equation}
We can thus expect the power law behaviors of $f(t)$ for $0<\alpha<1$ and $1<\alpha<2$.
This is confirmed by numerical simulation as shown in the next section.

\item For $\alpha=2$, the left hand side of Eq. (\ref{FP_powerlaw}) is a monotonically decreasing function, and therefore there exists one unstable fixed point as in the case of $\alpha>2$.
Nevertheless one may observe power-law behavior.
The time evolution equation reads
\begin{equation}
\label{FP_powerlaw_alpha2}
\dot{f} = \frac{f_{\min}^{-1}- f_{\max}^{-1}}{f^{-1}- f_{\max}^{-1}} f_0 - f.
\end{equation}
If we choose $f_0 = (f_{\min}^{-1}- f_{\max}^{-1})^{-1}$, the above equation reduces to 
\begin{equation}
\label{FP_powerlaw_alpha2_critical}
\dot{f} = \frac{f^2/ f_{\max}}{1- f/f_{\max}},
\end{equation}
which have a solution of $f\propto (t_f-t)^{-1}$ unless $f/f_{\max}\simeq 1$.
In this case, however, the power law behavior is observed even if the bifurcation does not occur.

\item For $\alpha=1$, the distribution reads
\begin{equation}
p(f) = \frac{1}{f \log\left(\frac{f_{\max}}{f_{\min}}\right)}.
\end{equation}
By computing $\Phi(f)$, the fixed points are given by the following equation.
\begin{equation}
f \log\left(\frac{f}{f_{\max}}\right) = f_0 \log\left(\frac{f_{\min}}{f_{\max}}\right).
\end{equation}
We can also show that the saddle-node bifurcation occurs at $f_0=f_{\max}/e\log(f_{\max}/f_{\min})$ and $f=e^{-1}f_{\max}$, and therefore we can expect power-law behaviors.
\end{itemize}


\section{Numerical results: Global load sharing model}
In the following, numerical results are shown for both the GLS and the LLS models. In each case, the system comprises $10^5$ fibers and the results are averaged over $10^4$ configurations. The main aim for the numerical study is to confirm the analytical results given in the previous section as well as to understand the time evolution of the system with more general threshold distributions. In case of the GLS model, we have used three different threshold distributions: uniform, Weibull and power laws. The numerical results are compared with the analytical ones. In case of the LLS model, the numerical results shown here are restricted to the uniform threshold distribution only.

For both the systems (GLS and LLS), we show the strain rate as a function of time in the system under a constant load, which is slightly above the critical value. Using the number of unbroken fibers after the $t^{th}$ redistribution (at time $t$), which is denoted by $L_t$, the strain $\epsilon(t)$ is represented as $F/L_t$ with the force $F$ applied to the system. This is indeed identical to the force per fiber $f$ in the GLS model.
However, in the LLS model, this strain may be interpreted as the average value (averaged over all surviving fibers), as the force per fiber is inhomogeneous. In the same manner, the strain rate is given as the time derivative of the average strain: $\dot{\epsilon}_t=F/L_{t+1}-F/L_t$. Again, this is identical to $\dot{f}$ in the GLS model, whereas it is the averaged value in the LLS model.


\subsection{Uniform threshold distribution: Comparison of analytical and numerical results}
\begin{figure}[ht]
\centering
\includegraphics[width=6cm, keepaspectratio]{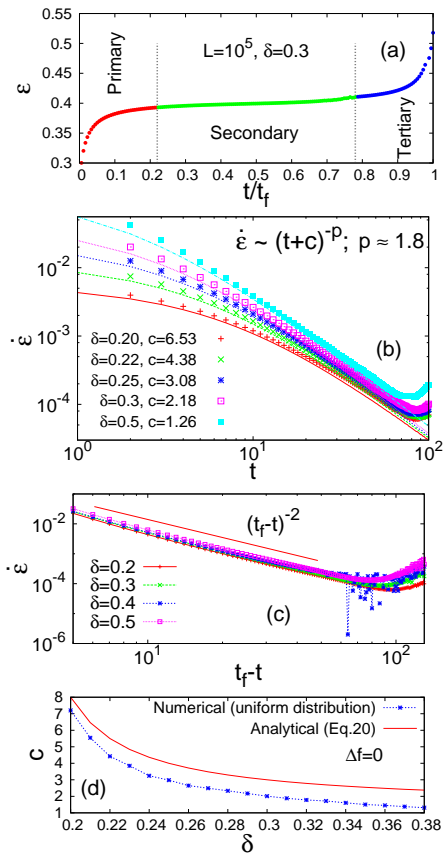}
\caption{(Color online) (a) Behavior of strain, showing the primary, secondary and tertiary stage in the time evolution normalized by failure time $t_f$. (b) \& (c) Variation of the strain rate with time respectively in the primary and tertiary stage, along with the comparison with the analytical findings (lines with no points). (d) Variation of c-value with $\delta$, the strength of disorder, in the primary stage.}
\label{Omori_Analytic}
\end{figure}

Here the threshold of each fiber is chosen from a uniform distribution defined on the interval of $[0.5-\delta, 0.5+\delta]$. Fig.\ref{Omori_Analytic}(a) shows the creep-like behavior observed under such critically stressed condition. The behavior shows all three stages: primary (red), secondary (green) and tertiary (blue). The strain rate at primary and tertiary stage is observed closely and shown in Fig.\ref{Omori_Analytic}(b) and Fig.\ref{Omori_Analytic}(c) respectively. 
The time evolution of strain rate in the primary stage (Fig.\ref{Omori_Analytic}b) follows the Omori-Utsu law and matches satisfactorily with the analytical expression (solid lines drawn according to Eq.\ref{MainOmori}). Also in the tertiary stage, the numerical results match with the inverse Omori law given by Eq.(\ref{MainInvOmori}). We will discuss this in details later in this paper. Also Fig.\ref{Omori_Analytic}d compares the analytical (see Eq.\ref{c-value}) and numerical $c$-values at different degrees of disorder, under the condition $f=f_c$. A probable reason for the discrepancies between analytical and numerical results will be the assumption made in Eq.(\ref{ref1}).


\subsection{Weibull distribution}
If the constituent fibers themselves are sufficiently macroscopic objects, the strength of a single fiber may obey extreme statistics.
In this case one can consider the Weibull distribution for the fiber strength.
\begin{equation}
\int_f ^{\infty} p(y)dy = \exp\left[ -\left(\frac{f}{\widetilde{f}}\right)^{\beta}\right],
\end{equation}
where $\beta > 0$ and $\widetilde{f}$ is a constant.
This leads to the time evolution equation:
\begin{equation}
\label{ODE_Weibull}
\tau\dot{x}=x_0 \exp\left( x^{\beta}\right) -x,
\end{equation}
where $x=f/\widetilde{f}$.

Whereas the uniform distribution case is controlled by the only one dimensionless parameter, the Weibull distribution case involves two dimensionless parameter, $\beta$ and $x_0$. Note that very large or very small $\beta$ values correspond to less heterogeneity and the intermediate values of $\beta$ may represent disordered systems. However, the parameter $\beta$ does not affect the qualitative behavior of the system as shown below.


\subsubsection{Saddle-node bifurcation}      
First, we show saddle-node bifurcation also occurs for the Weibull distribution case irrespective of the value of $\beta$.
The fixed points of Eq. (\ref{ODE_Weibull}) must satisfy
\begin{equation}
x_0 \exp\left( x^{\beta}\right) = x.
\end{equation}
Since $x>0$, it is more convenient to take the logarithm.
\begin{equation}
x^{\beta}-\log x = -\log x_0.
\end{equation}
As the left-hand side is a simple concave function of $x$ for positive $\beta$, the above equation must have two solutions at sufficiently small $x_0$.
One can show that the smaller solution is the stable fixed point, whereas the larger one is unstable.
At a critical value of $x_0 = (e\beta)^{-1/\beta}$, these two fixed points merge and disappear.
This is the saddle-node bifurcation as in the case of the uniform threshold distribution.
For $x_0 > (e\beta)^{-1/\beta}$, there is no fixed point and the force per fiber increases rapidly with time.
Therefore, we can expect the power-law behaviors with the exponent $2$ as discussed in the previous section.


\subsubsection{Dependence on disorder $\beta$}
To check the response of the model to the disorder introduced, we have studied the variation of strain rate ($\dot{\epsilon}$) with time at different disorder values while the applied stress is kept constant at the critical value.
\begin{figure}[ht]
\centering
\includegraphics[width=6cm, keepaspectratio]{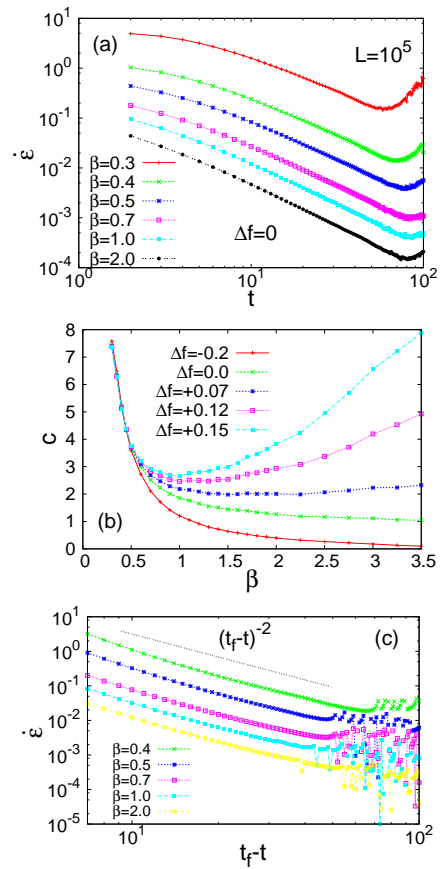}
\caption{(Color online) Time evolution of the model with Weibull threshold distribution. (a) Omori-like behavior in the primary stage with a continuous variation of disorder. (b) Variation of $c$-value with disorder at different loading conditions. (c) Inverse Omori like behavior close to the failure point.}
\label{Omori_Numerical_Disorder}
\end{figure} 
Figure \ref{Omori_Numerical_Disorder} shows this $\dot{\epsilon}$ v/s $t$ behavior with a continuous change in $\beta$ (and hence change in disorder).
Here also we observe the Omori and inverse Omori laws in the primary and tertiary stages, respectively.
\begin{itemize}
\item Primary stage : $\dot{\epsilon}=\displaystyle\frac{k}{(t+c)^p}$, $p\approx1.8$. \vspace{-0.07cm}
\item Tertiary stage : $\dot{\epsilon}=\displaystyle\frac{k^{\prime}}{(t_f-t)^{p^{\prime}}}$, $p^{\prime}\approx2.0$.
\end{itemize}    
where $t_f$ is the time of breakdown of the system. 

Both exponents $p$ and $p^{\prime}$ show satisfactory match with the analytical results. In the primary stage, the value of $c$ changes with a continuous variation of $\beta$. The applied stress is kept constant here at the critical value. Strain rate produced in the model at critical stress and corresponding to different $\beta$ values (see Fig.\ref{Omori_Numerical_Disorder}a) in the primary stage are fitted with the Omori law for different $c$ values. The variation of $c$-value with $\beta$ is shown in Fig.\ref{Omori_Numerical_Disorder}(b) at different loading condition. $\Delta f$ shows the deviation in applied stress from the critical value $f_c$. $\Delta f=0$ corresponds to the critically loading condition. A positive $\Delta f$ tells us that the model is overloaded, while a negative value of $\Delta f$ leads to situation where the applied load is less than the critical one. $c$ attains a higher value at both low and high $\beta$ value and hence at low disorder limit. This non-monotonic behavior is very prominent where the system is more and more overloaded ($\Delta f>0$). For $\Delta f \approx0$ or $\Delta f<0$, we have to go to relatively higher value of $\beta$ to observe this increment in $c$-value. Finally Fig.\ref{Omori_Numerical_Disorder}(c) shows the strain rate, close to the failure point. When $t$ approaches $t_f$, $\dot{\epsilon}$ increases in a scale-free manner with an exponent $-2$, independent of the disorder introduced in the model. Also, as discussed before, we have a zero c-value here.


\subsubsection{Dependence on the applied stress}      
Next we have investigated the effect of applied stress more closely focusing on the primary stage only.
\begin{figure}[ht]
\centering
\includegraphics[width=6cm, keepaspectratio]{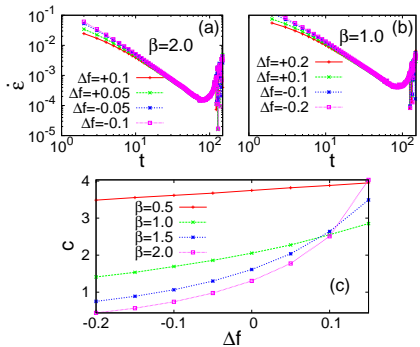}
\caption{(Color online) Omori like behavior at different loading condition. (a) \& (b) $\dot{\epsilon}$ v/s $t$ for two different disorder values $\beta=2.0$ and $\beta=1.0$. (c) Variation of $c$-value with $\Delta f$ for $0.5<\beta<2.0$.}
\label{Omori_Numerical_Sigma}
\end{figure} 
Fig.\ref{Omori_Numerical_Sigma}a and Fig.\ref{Omori_Numerical_Sigma}b shows the Omori like behavior under different loading conditions. At high $\beta$ value the system responses quite well with varying applied stress. On the other hand at low $\beta$ value, c changes very slowly with $\Delta f$. These different responses can be expressed through a continuous variation of $c$-value with $\Delta f$. Fig.\ref{Omori_Numerical_Sigma}c shows the $c$-value v/s $\Delta f$ variation at different $\beta$ (hence at different disorder values). As previously discussed, at low $\beta$, $c$ starts with a relatively higher value and gradually increases with $\Delta f$. For higher $\beta$ value, $c$ attains a lower value initially but increases rather faster with $\Delta f$. Hence, the rate of change of such c-value is relatively higher for high disorder values.


\subsubsection{Variation of c-value on $\Delta f-\delta$ plane}     
\begin{figure}[ht]
\centering
\includegraphics[width=6cm, keepaspectratio]{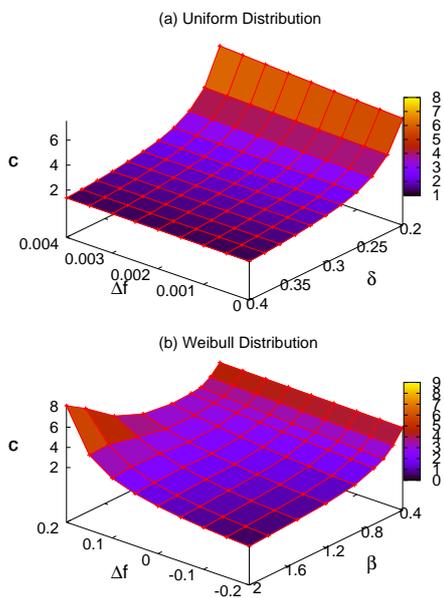}
\caption{(Color online) Variation of $c$-value when both $\Delta f$ and $\beta$ are continuously varying parameters. Results are shown for (a) Uniform and (b) Weibull distribution.}
\label{3D_Diagram}
\end{figure}

Finally we have reached a point where we can explain the behavior of the $c$-value with respect to the parameters $\beta$ and $\Delta f$.
In Fig.\ref{3D_Diagram}a and \ref{3D_Diagram}b we have shown the $c$-value in the primary stage as function of both disorder and applied stress in case of the uniform and the Weibull distributions.
The value of $c$ is higher at less disorder (low $\delta$ for uniform distribution and very low or very high $\beta$ for Weibull distribution) and gradually deceases when we go to higher disorder. At any particular $\delta$ or $\beta$, the $c$-value increases with increasing $\Delta f$. \\


\subsection{Power-law threshold distribution}
Next we have carried out the numerical simulation where the thresholds are chosen randomly from a power law distribution with a variable exponent.
Following the analytical results, we introduce some cut-off values for the distribution depending on the value of $\alpha$. Irrespective of the value of $\alpha$, we must choose a sufficiently large width that remains constant throughout the numerical simulation. Here it is chosen to be [1,$10^3$], where the exponent $\alpha$ varies between $0.5$ and $2.5$. 
\begin{figure}[ht]
\centering
\includegraphics[width=6cm, keepaspectratio]{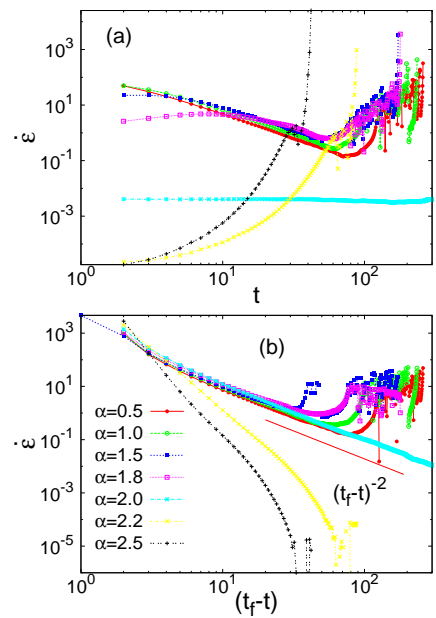}
\caption{(Color online) Strain rate vs time in the (a) primary and (b) tertiary stage of the creep process. The threshold distribution considered here is $p(f)=f^{-\alpha}$ within the window [1,$10^3$].}
\label{Omori_Powerlaw_Initial}
\end{figure} 
This variation in the exponent essentially covers all the three situations discussed in the analytical study (see \ref{Powerlaw_Analytical}).

Fig.\ref{Omori_Powerlaw_Initial} illustrates the behaviors of strain rate in the primary and the tertiary stages, respectively, where the qualitative difference is apparent for different values of the exponent $\alpha$:
\begin{enumerate}[label=(\roman*)]
\item For $\alpha<2$, we obtain a power law decrease of strain rate $\dot{\epsilon}$ in the primary stage. At the same time in the tertiary stage $\dot{\epsilon}$ increases obeying the inverse Omori law until it reaches global failure. The $p$ values at primary and tertiary stages are respectively $1.8$ and $2.0$. 
\item For $\alpha>2$, the model shows brittle response. In this limit $\dot{\epsilon}$ increases exponentially in the primary stage and reaches to global failure much more rapidly and does not exhibit the inverse Omori law in the tertiary stage.      
\item At $\alpha = 2$, the power law behavior is observed only in the form of inverse Omori law while the strain rate is almost constant in the primary regime.       
\end{enumerate}
All these behaviors are consistent with the analytical results.


\section{Numerical results: Local load sharing model}
To include the effect of local stress concentration, we assume that the close neighborhood of a broken fiber is affected much more than the other parts of the model. For this purpose, we redistribute the stress of a broken fiber over a finite distance, known as the stress release range $R$.
A recent study has already shown that there exits the critical range value $R_c$, above which the model shifts to the mean-field limit.
The critical value depends on the system size $L$ as $R_c \sim L^{2/3}$ \cite{Biswas}. In this paper, instead of $R$, we have used $\rho=R/R_c$ as the parameter, and thus $\rho\ge1$ corresponds to the mean field limit of the model. Here the uniform threshold distribution is adopted and therefore we investigate the effects of disorder by changing $\delta$, as well as the effects of the interaction range by changing $\rho$. Again, the external load remains to be slightly above the critical value. 


\subsubsection{Role of disorder}  
\begin{figure}[ht]
\centering
\includegraphics[width=6cm, keepaspectratio]{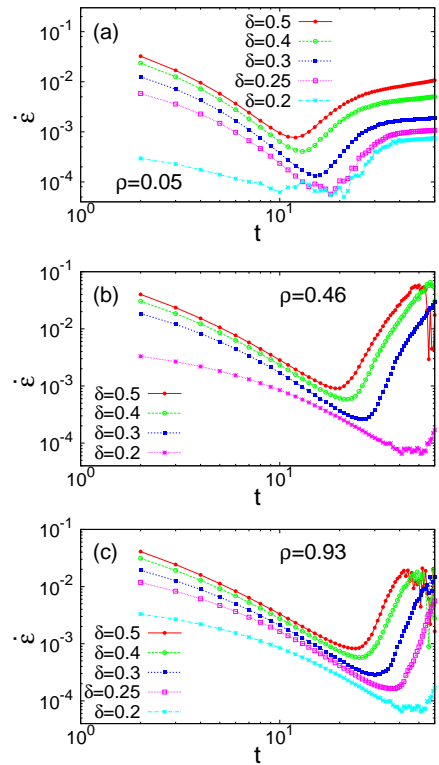}
\caption{(Color online) Variation of strain rate with time (in the primary stage) for disorder $\delta$, ranging in between $0.2$ and $0.5$, while a critical stress is applied on it. The study is repeated for $\rho=0.93$, $0.46$ and $0.05$.}
\label{Omori_Disorder1}
\end{figure}

Figure \ref{Omori_Disorder1} shows the behavior of strain rate in the primary stage with different values of local stress concentration parameter $\rho$: $0.93$, $0.46$ and $0.05$.
In case of $\rho=0.05$, the stress is redistributed up to a small range, whereas $\rho=0.93$ is close to the mean-field limit. As a result, it is expected to obtain the Omori-Utsu law in the primary stage when the $\rho$ is close to 1 (see Fig.\ref{Omori_Disorder1}c). Interestingly, we observe this behavior to sustain even for lower $\rho$, namely more stress localization. The $c$-value also changes with disorder in this limit.
The extra feature that we get with stress localization is a varying exponent $p$ in the Omori-Utsu law.


\subsubsection{Role of stress release range}
Here we have studied the model at a constant disorder but with varying stress release range. By changing a variable $\rho$ the model shifts from the mean-field limit to another limit where stress redistribution is extremely localized.  
\begin{figure}[ht]
\centering
\includegraphics[width=6cm, keepaspectratio]{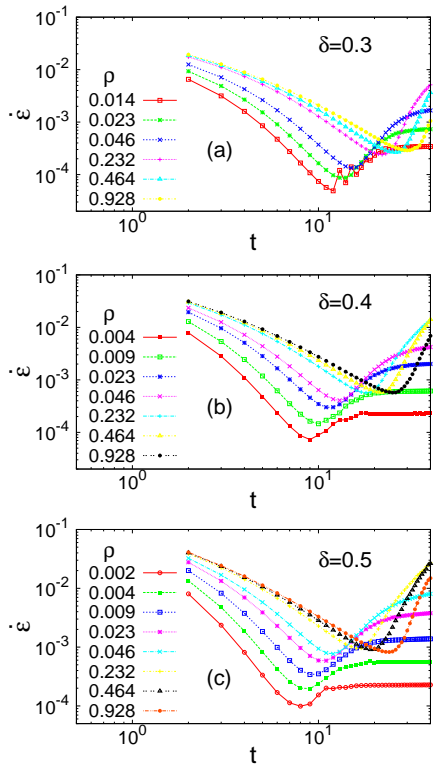}
\caption{(Color online) Plot for $\dot{\epsilon}$ vs $t$ in the primary stage at three different disorder values ($\delta=0.3$, 0.4 and 0.5). The stress is kept constant at the critical value while $\rho$ is continuously varied.}
\label{Omori_Range}
\end{figure} 
Figure \ref{Omori_Range} shows the time evolution of the strain rate with several values of $\rho$. The study is repeated for three different disorder values. The slope in the Omori like behavior clearly shows an increment while $\rho$ is decreased. Also at very low $\rho$, the exponent $p$ changes with disorder. This variation of $p$ with disorder was absent in the mean field limit.

  
\subsubsection{Variation of c-value and exponent p}
To understand the effect of such stress localization, we have studied the $c$-value and the exponent $p$ with a continuous variation of stress release range $\rho$ between 0 and 1.5. As we have already mentioned, for $\rho\ge1$ the model enters its mean-field limit.
  
\begin{figure}[ht]
\centering
\includegraphics[width=6cm, keepaspectratio]{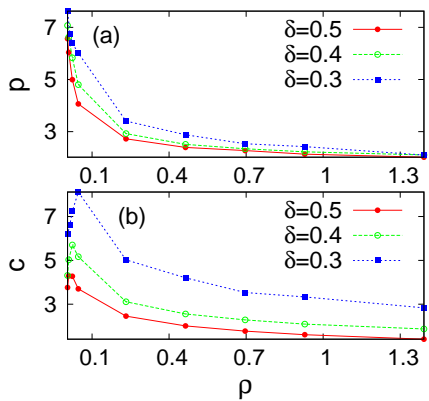}
\caption{(Color online) Variation of the $c$-value and and the exponent $p$ with increasing stress localization. The model approaches the mean-field limit toward $\rho=1$, and $p$ approaches its mean-field value $1.8$ gradually with increasing $\rho$.}
\label{Exponents_Range1}
\end{figure} 
For smaller interaction range ($\rho <1$), both $p$ and $c$ take large values. As $\rho$ is increased, these two quantities decrease gradually.
Throughout the stage $0<\rho\le1.5$, the $c$-value remains a function of disorder $\delta$ and increases as we go to lower $\delta$ values.
So, $c$ can be expressed as follows: $c=\Phi(\rho,\delta)$ for $0<\rho\le1.5$, where $\Phi$ is an decreasing function of both stress release range $\rho$ and disorder $\delta$. On the other hand, the exponent of $p$ is function of both disorder $\delta$ and stress localization $\rho$ for $\rho<1.0$. For $\rho>1.0$, $p$ takes a value 1.8 independent of $\delta$ and $\rho$, which is the mean field exponent for Omori law we obtained previously.


The results in the mean field limit are already shown for three different distributions: uniform, Weibull and power law. An universal behavior is observed in the time evolution of strain rate (or force per fiber) for all these three distributions. With local stress concentration, we have shown the results with uniform distribution. The universality of these results are also checked with a Weibull distribution with shape parameter $\beta$ and a power-law distribution with exponent $-1$ ranging from $10^{-\eta}$ to $10^{\eta}$. The parameters $\eta$ and $\beta$ controls the disorder here. 


\section{Discussions}
Here we discuss the relevance of the Omori-Utsu and the inverse Omori laws in a more general context. Although our results are presented in terms of the strain rate, they should apply to more general cases if there is a relation between the rate of micro-fracture events and the strain rate, such as $n(t) \propto \dot{\epsilon}(t)^q$, where $n(t)$ is the rate of micro-fracture events and $q$ a positive exponent.
As the rupture of a single fiber may correspond to a single micro-fracture event in the present system, $n(t) \propto \dot{\epsilon}(t) L^2(t)$, where $L(t)$ is the number of remaining fibers at time $t$.

Noting that aftershocks are caused by the abrupt stress change caused by a mainshock, the algorithm adopted here, in which a finite stress is applied to the system at $t=0$, may model such a stress change caused by a mainshock. In this sense, the Omori-Utsu law obtained in the present model mimics the dynamics after a mainshock for earthquakes to some extent. We obtain the exponent $p\simeq 2$ in the mean-field model irrespective of the other details such as the threshold distribution, whereas $p$ ranges from $0.6$ to $0.8$ in creep test and from $0.7$ to $1.6$ for earthquakes. The difference is significant but the quantitative agreement is not necessarily here because of the simplicity of the mean-field model. In contrast, the difference is even larger for the LLS model. Noting that the LLS model is generally more unstable than the GLS model, we may obtain smaller exponent for more stable systems. For instance, introducing a probabilistic rule for the elementary fracture process might lead to smaller exponent because it can inhibit the cascade-like instability of fracture caused by the load redistribution to slow down the relaxation.

The $c$-value is a characteristic time for the power-law relaxation that results from the abrupt stress loading and therefore it is regarded as a relaxation time for stress. The elementary stress relaxation time in our model is $\tau$, which makes a single time step. It is indeed the only intrinsic time constant in our model and therefore the $c$-value should be scaled with $\tau$ from dimensional analysis. The $c$-value is thus mostly dominated by the nature of $\tau$. For instance, if the stress relaxation time depends on the total load $F$, the analysis given in this study still applies and yields the load-dependent $c$-value.

In the GLS model, the analytical expression for the $c$-value is obtained for a class of threshold distributions. Apart from the trivial dependence on $\tau$, the $c$-value depends on three parameters: $a$, $f_c$, and $f_0$. Among them, $a$ and $f_c$ are determined mostly by the threshold distribution via Eqs. (\ref{Phi}) and (\ref{expansion2}) but they also depend on $f_0$ because $\Phi(f)$ is proportional to $f_0$. Therefore, $a$ should be proportional to $f_0$. Noting that $f_0$ is proportional to the total load $F$ as $f_0 = F/L_0$, Eq. (\ref{c-value_general}) implies the load dependence of the $c$-value. Although the $c$-value is found to increase with the load in this study, in view of Eq. (\ref{c-value_general}), it can be a decreasing function of the external load if $f_c > 2 f_0$. This actually means $f_c > 2f_0^*$ and therefore it depends on the threshold distribution. This condition is not satisfied for the distribution functions investigated here and hence the $c$-value exhibits only positive dependence on the external load.

\section{Conclusions}
We have studied the time evolution of fiber bundle model under a constant external load being slightly above the critical value with some variations in the load redistribution process: the global-load sharing and the local-load-sharing models. The strain rate in the primary and the tertiary stages follows the Omori-Utsu and the inverse Omori laws respectively. In the local-load-sharing model, both the exponent $p$ and the $c$-value are decreasing functions of disorder and the interaction range. Above a certain stress release range ($\rho>1$), the local-load-sharing model exhibits essentially the same behavior as that of the mean filed limit; namely, the exponent for the Omori-Utsu law attains a constant value ($\approx 1.8$) and $c$ is still a decreasing function of disorder. Despite the simplicity of the model and the absence of any thermal activation process, the system exhibits creep-like behaviors with all the three stages: primary, secondary and tertiary. This in turn implies that the probabilistic rule is not essential for a power-law behavior in creep deformation.




\begin{thebibliography}{99}
\bibitem{OmoriMain1} F. Omori, J. Coll. Sci., Imp. Univ. Tokyo {\bf 7}, 111 (1894).
\bibitem{Utsu} T. Utsu, J. Fac. Sci. Hokkaido Univ. Ser. VII {\bf 3}, 379 (1965).
\bibitem{GR} B. Gutenberg and C. F. Richter, Seismicity of the Earth and Associated Phenomena (Princeton University Press, Princeton, NJ, 1954);,
Bull. Seismol. Soc. Am. {\bf 46}, 105 (1954).
\bibitem{OmoriReview} T. Utsu, Y. Ogata, and R. S. Matsuura, J. Phys. Earth {\bf 43}, 1 (1995).
\bibitem{Omori1} R. Shcherbakov, D. L. Turcotte and J. B. Rundle, Geophys. Res. Lett., Volume {\bf 31}, Issue 11 (2004).
\bibitem{Peng} Z. G. Peng, J. E. Vidale, H. Houston, Geophys. Res. Lett. {\bf 33}, doi:10.1029/2006GL026744 (2006).
\bibitem{Enescu} B. Enescu, J. Mori, M. Miyasawa, J. Geophys. Res. {\bf 112}, B04310, doi:10.1029/2006JB004629 (2007).
\bibitem{Narteau} C. Narteau, S. Byrdina, P. Shebalin, D. Schorlemmer, Nature {\bf 462}, 642 (2009).
\bibitem{Scholz} C. H. Scholz, Bull. Seism. Soc. Am. {\bf 58}, 399 (1968).
\bibitem{Hirata} T. Hirata, J. Geophys. Res. {\bf 92}, 6215 (1987).
\bibitem{Schubnel} A. Schubnel, B. D. Thompson, J. Fortin, Y. Gu\'eguen, and R. P. Young, Geophys. Res. Lett., {\bf 34}, L19307 (2007).
\bibitem{Andrade} E.N. Da C. Andrade, Proc. R. Soc. London {\bf A 84}, 1 (1910).
\bibitem{Bouchon} M. Bouchon, V. Durand, D. Marsan, H. Karabulut, and J. Schmittbuhl, Nat. Geosci {\bf 6}, 299 (2013).
\bibitem{Daniels}H. E. Daniels and T. H. R. Skyrme, Adv. Appl. Probab. {\bf 21}, 315 (1989).
\bibitem{RevModPhys82} S. Pradhan, A. Hansen, and B. K. Chakrabarti, Rev. Mod. Phys. \textbf{82}, 499 (2010).
\bibitem{Book} A. Hansen, P. C. Hemmer \& S. Pradhan, \textit{The Fiber Bundle Model: Modeling Failure in Materials} Wiley VCH Berlin (2015).
\bibitem{Ciliberto} S. Ciliberto, A. Guarino, R. Scorretti, Physica D {\bf 158} (1–4), 83 (2001).
\bibitem{Pradhan2003} S. Pradhan and B. K. Chakrabarti, Int. J. Mod. Phys. B {\bf 17}, 5565 (2003).
\bibitem{Shcherbakov} R. Shcherbakov and D. L. Turcotte, Theor. Appl. Fract. Mech. {\bf 39}, 245 (2003).
\bibitem{Ben-Zion} Y. Ben-Zion and V. A. Lyakhovsky, Geophys. J. Int. {\bf 165}, 197 (2006).
\bibitem{Saichev} A. Saichev and D. Sornette, Phys. Rev. E {\bf 71}, 016608 (2005).
\bibitem{Danku} Z. Danku and F. Kun, Sci. Rep. {\bf 3}, 2688 (2013).
\bibitem{Pradhan2007} S. Pradhan and P. C. Hemmer, Phys. Rev. E {\bf 75}, 056112 (2007).
\bibitem{Pierce} F. T. Pierce, J. Text. Ind. {\bf 17}, 355 (1926). 
\bibitem{Phoenix} S. L. Phoenix, Adv. Appl. Probab. {\bf 11}, 153 (1979). 
\bibitem{Smith} R. L. Smith and S. L. Phoenix, J. Appl. Mech. {\bf 48}, 75 (1981). 
\bibitem{Newman} W. I. Newman and S. L. Phoenix, Phys. Rev. E {\bf 63}, 021507 (2001). 
\bibitem{Harlow2} D. G. Harlow and S. L. Phoenix, J. Compos. Mater. {\bf 12}, 314 (1978).
\bibitem{Harlow3} D. G. Harlow and S. L. Phoenix, Adv. Appl. probab. {\bf 14}, 68 (1982).
\bibitem{Smith2} R. L. Smith, Proc. R. Soc. London, Ser. A {\bf 382}, 179 (1982).
\bibitem{sroyepl} S. Roy and P. Ray, Europhys. Lett. {\bf 112}, 26004 (2015).
\bibitem{llssystemsize1} Gomez et al. Phys. Rev. Lett. {\bf 71}, 380 (1993).
\bibitem{sroy} S. Roy, Phys. Rev. E. {\bf 96}, 042142 (2017).
\bibitem{Biswas} S. Biswas, S. Roy and P. Ray, Phys. Rev. E {\bf 91}, 050105(R) (2015).
\end{thebibliography}
\end{document}